\documentstyle[12pt]{article}

\begin{document}

\baselineskip=7mm

\newcommand{\TeV}{\,{\rm TeV}}
\newcommand{\GeV}{\,{\rm GeV}}
\newcommand{\MeV}{\,{\rm MeV}}
\newcommand{\keV}{\,{\rm keV}}
\newcommand{\eV}{\,{\rm eV}}
\newcommand{\Tr}{{\rm Tr}\!}
\renewcommand{\arraystretch}{1.2}
\newcommand{\be}{\begin{equation}}
\newcommand{\ee}{\end{equation}}
\newcommand{\bea}{\begin{eqnarray}}
\newcommand{\eea}{\end{eqnarray}}
\newcommand{\ba}{\begin{array}}
\newcommand{\ea}{\end{array}}
\newcommand{\bmat}{\left(\ba}
\newcommand{\emat}{\ea\right)}
\newcommand{\refs}[1]{(\ref{#1})}
\newcommand{\ler}{\stackrel{\scriptstyle <}{\scriptstyle\sim}}
\newcommand{\ger}{\stackrel{\scriptstyle >}{\scriptstyle\sim}}
\newcommand{\lag}{\langle}
\newcommand{\rag}{\rangle}
\newcommand{\ns}{\normalsize}
\newcommand{\cm}{{\cal M}}
\newcommand{\gr}{m_{3/2}}
\newcommand{\p}{\partial}

\def\321{$SU(3)\times SU(2)\times U(1)$}
\def\tl{{\tilde{l}}}
\def\tL{{\tilde{L}}}
\def\bd{{\overline{d}}}
\def\tL{{\tilde{L}}}
\def\a{\alpha}
\def\b{\beta}
\def\g{\gamma}
\def\c{\chi}
\def\d{\delta}
\def\D{\Delta}
\def\db{{\overline{\delta}}}
\def\Db{{\overline{\Delta}}}
\def\e{\epsilon}
\def\l{\lambda}
\def\n{\nu}
\def\m{\mu}
\def\nt{{\tilde{\nu}}}
\def\p{\phi}
\def\P{\Phi}
\def\x{\xi}
\def\r{\rho}
\def\s{\sigma}
\def\t{\tau}
\def\th{\theta}
\def\ne{\nu_e}
\def\nm{\nu_{\mu}}

\renewcommand{\Huge}{\Large}
\renewcommand{\LARGE}{\Large}
\renewcommand{\Large}{\large}

\begin{titlepage}
\title{\bf Quasi Goldstone Fermion As a Sterile Neutrino\\
                          \vspace{-4cm}
                          \hfill{\ns IC/95/164\\}
                          \hfill{\ns PRL-TH/95-11\\}
                          \hfill{\ns hep-ph/9507371\\[.3cm]}
                          \vspace{3cm} }

\author{ Eung Jin Chun$^\dagger$ \hspace{.4cm}
         Anjan S.~Joshipura$^*$ \hspace{.4cm}
         Alexei Yu.~Smirnov$^{\dagger, \#}$ \\[.5cm]
  {\ns\it $^\dagger$International Center for Theoretical Physics}\\
  {\ns\it P.~O.~Box 586, 34100 Trieste, Italy} \\[.3cm]
  {\ns\it $^*$Theoretical Physics Group, Physical Research Laboratory}\\
  {\ns\it Navarangpura, Ahmedabad, 380 009, India} \\[.3cm]
  {\ns\it $^\#$Institute for Nuclear Research, Russian Academy of Sciences}\\
  {\ns\it 117312 Moscow, Russia} }
\date{}
\maketitle
\vspace{2cm}
\begin{abstract} \baselineskip=7mm {\normalsize
The existence of sterile neutrino is hinted by simultaneous explanation of
diverse neutrino anomalies.
We suggest that the quasi Goldstone fermions (QGF) arising in
supersymmetric theory as a result of spontaneous breaking of
global symmetry like the Peccei-Quinn symmetry or the lepton number symmetry
can play a role of the sterile neutrino.
The smallness of mass of QGF ($m_S \sim 10^{-3}-10$ eV) can be
related to the specific choice of superpotential or K\"ahler potential
(e.g., no-scale kinetic terms for certain superfields).
Mixing of QGF with neutrinos implies the $R$-parity violation.
It can proceed via the  coupling of QGF with the Higgs
supermultiplets or directly with the lepton doublet.
A model which accounts for the solar and atmospheric anomalies and the
dark matter is presented.
%
}\end{abstract}

\thispagestyle{empty}
\end{titlepage}

\section{Introduction}

All the experimentally known fermions transform non-trivially
under the gauge group \321 of the standard model (SM).
However there are  experimental
hints in the neutrino sector which suggest the existence of \321
- singlet fermions mixing appreciably with the known neutrinos.
These hints come from
(a) the deficits in the solar \cite{solar} and atmospheric \cite{atm}
neutrino fluxes
(b) possible need of significant hot component \cite{dm} in the dark
matter of the universe  and
(c) some indication of $\bar{\nu}_e-\bar{\nu}_{\mu}$ oscillations in
the laboratory \cite{lsnd}.
These hints can be reconciled with each
other if there exists a fourth very light ($< {\cal O}$(eV))
neutrino mixed with some of the known  neutrinos preferably
with the electron one.  The fourth neutrino is
required to be sterile in view of the strong bounds on number of
neutrino flavours coming both from the LEP experiment
as well as from the primordial nucleosynthesis \cite{ns}.

The existence of very light sterile neutrino demands theoretical
justification since unlike the active neutrinos, the mass of the
sterile state is not protected by the gauge symmetry of the SM
and hence could be very large.
Usually the sterile neutrino is considered on the same footing as the active
neutrinos and some ad hoc symmetry is introduced to keep this neutrino light.
Recently there are several attempts to construct models for sterile
neutrinos which have the origin  beyond the usual lepton
structure \cite{paper1,mirror,ma}. In particular in Ref.~\cite{paper1}
we suggested a possibility that supersymmetry (SUSY) may be responsible
for both the existence and the lightness of the sterile fermions.

One could consider three different ways in which supersymmetry can
keep sterile states very light.

\noindent (1) Combination of supersymmetry and the (continuous)
$R$ symmetry present in
many supersymmetric models may not allow a mass term for the
light sterile state.

\noindent (2) Spontaneous breakdown of some other global symmetry in
supersymmetric theory can lead to massless fermions which form
the superpartners of the Goldstone bosons.

\noindent (3) The spontaneous breakdown of the global supersymmetry
itself would give rise to a massless fermion, the goldstino.

The mechanism (1) and its phenomenological consequences
were discussed in Ref.~\cite{paper1}. Mechanism (3) though appealing
is not favoured phenomenologically in view of the difficulties
in building realistic models based on the spontaneously broken
global SUSY. We discuss in this paper implications
of the  mechanism (2) concentrating for definiteness on the simplest
case of a global $U(1)_G$.

The spontaneously broken global symmetries are required
for reasons unrelated to the existence of light sterile states.
The most interesting examples being spontaneously broken lepton number
symmetry \cite{cmp} and the Peccei-Quinn (PQ) symmetry imposed \cite{pq}
to solve the strong CP problem.
The PQ symmetry arise naturally in many supersymmetric models.
Apart from solving the strong CP problem, this symmetry
can also explain the smallness of the $\m$-parameter \cite{mu,chun3}.
Phenomenologically consistent breaking of these symmetries generally
needs \cite{dfs} Higgs fields which are singlets of \321.
In the supersymmetric context this
automatically generates massless sterile fermion. While the
existence of these quasi Goldstone fermions (QGF) is logically
independent of neutrino physics, there are good
reasons to expect that these fermions will couple to  neutrinos.
Indeed, in the case of lepton number symmetry
the superfield which is mainly responsible
for the breakdown of $U(1)_L$ carries nontrivial $U(1)_L$-charge and
therefore it can directly couple to leptons if the charge is appropriate.
In the case of the PQ symmetry, $U(1)_{PQ}$,
this superfield  could couple to the Higgs supermultiplet.
If theory contains small violation of $R$ parity then this mixing
with Higgs gets communicated to the neutrino sector.
Thus the occurrence of the QGF can have implications for neutrino
physics. We wish to discuss in this paper prospects for building realistic
models based on this mechanism.

In the following section we elaborate upon  the expected properties
of the QGF, especially their masses when SUSY is broken.
Section 3 discusses various mechanisms of mixing
of these  fermions with the active neutrinos.
Explicit model based on the scenario presented
in section 2 and 3 is given in section 4 and the last section
presents our conclusions.

\section{Quasi Goldstone fermions and their masses}

In this section and subsequently, we will consider the following
general superpotential
\be \label{w}
W=W_{MSSM}+W_S+W_{mixing} \;,  \ee
where $W$ is assumed to be invariant under some global symmetry
$U(1)_G$. As we outlined in the introduction,
this symmetry may be identified with the
PQ symmetry, lepton number symmetry or combination thereof. The
first term in Eq.~\refs{w} refers to the superpotential of the minimal
supersymmetric standard model (MSSM).
The second term contains \321 singlet superfields which are responsible
for the breakdown of $U(1)_G$.
The minimal choice for $W_S$ is
\be \label{ws} W_S=\lambda (\s \s'- f_G^2) y \;, \ee
where $\s,\s'$ carry non trivial $G$-charges and $f_G$ sets the
scale of $U(1)_G$ breaking.
The last term of Eq.~\refs{w} describes mixing of the singlet
fields with the superfields of the MSSM.

In the supersymmetric limit the fermionic component of the Goldstone boson
is massless.  In the case \refs{ws} this Goldstone fermion
is contained in
\be \label{qgf} S=\frac{1}{\sqrt{2}}(\s-\s') \;. \ee
However, SUSY breakdown results in generation of mass of the
Goldstone fermion.  In general, this mass can be as big as SUSY breaking
scale, $m_{SUSY}$.  Broken supersymmetry itself cannot automatically protect
the masses of QGF in Eq.~\refs{qgf} much below $m_{SUSY}$.
In fact, the mass of QGF depends on the manner in which SUSY is broken and on
the way how this breaking is communicated to the singlet $S$.   It also
depends on the structure of superpotential and the scale $f_G$.
In the below we identify theories which
can allow for very light QGF ($m_S < 1$ eV).
As the case of special interest we will consider the mass of QGF and its
mixing with the electron neutrino:
\bea \label{parameters}
 m_S &\simeq& (2-3)\cdot 10^{-3} \eV \nonumber\\
 \sin\th_{es} &\simeq& \tan\th_{es} \simeq (2-6)\cdot 10^{-2} \;.
\eea
These values of parameters allow one to solve the solar neutrino
problem through the resonance conversion $\n_e \to S$ \cite{msw}.
\bigskip

One could consider different mechanisms for the QGF mass generation.\\
Let us note that in models with spontaneously broken global SUSY
the QGF generically acquire a mass of
${\cal O}(\frac {m_{SUSY}^2}{f_G})$ \cite{tam}.
But it can remain massless in spite of SUSY breaking
(a) if  SUSY is broken by a D-term of the gauge
field or (b) if the F-terms that break SUSY do not carry any G-charges.
The latter is exemplified by a simple generalization of Eq.~\refs{ws}:
$$ W_S=\lambda_1 (\s \s'- f_1^2) y_1 +\lambda_2 (\s \s'- f_2^2)y_2\;. $$
SUSY is broken in this example if $f_1^2\neq f_2^2$.
For a minimum with the F-terms: $F_{\s}=F_{\s'}=0$, the
Goldstone fermion in Eq.~\refs{qgf}
remains massless at the tree level in spite of the SUSY breakdown.
As we noticed before this version has phenomenological problems and further
on we will concentrate on possibilities related to supergravity.

The mass of the QGF in supergravity theory is typically of the
order of gravitino mass  $\gr$ ($=m_{SUSY}$) \cite{chun1,chun2,yam}.
For instance, the superpotential in Eq.~\refs{ws} leads to $m_S \sim \gr$
when generic soft terms of SUSY breakdown are allowed \cite{chun1}.
Howerver, the mass $m_S$ can be much smaller for specific choices of
1) the superpotential and/or 2) soft SUSY breaking terms. Let us consider
these possibilities in order.\\
1). The superpotential
$$ \lambda (\s \s'- X^2) y +\lambda'(X-f_G)^3 $$
is shown \cite{chun2} to generate the tree level mass
\be \label{small} m_S \sim  {\gr^2 \over f_G}  \ee
as in the global case if the minimal kinetic terms
of the fields are assumed.
For commonly accepted value of the PQ symmetry breaking scale, $f_G = f_{PQ}
= 10^{10}-10^{12}$ GeV, one gets from Eq.~\refs{small}
$m_S \sim (10-10^3)$ eV.
On the other hand, the value of $m_S$ in Eq.~\refs{parameters}
desired for explanation of the solar neutrino deficit requires
$f_G \sim 10^{16}$ GeV which can be related to the
grand unification scale.  To identify $f_G$ with $f_{PQ}$, one should overcome
the cosmological bound $f_{PQ}< 10^{12}$ GeV.
The bound can be removed by axion mixing with some other Goldstone boson in
their kinetic terms \cite{bbs} or by dilaton field
driven to small values in inflationary period  \cite{dvali}.
In this case however, the axion cannot play the role of cold dark matter.

2). Another possibility to get very light $S$ is based on the idea of
no-scale supergravity~\cite{noscale}.
The K\"ahler potential and the superpotential can be arranged
in such a way that supersymmetry breaking is communicated to the singlet $S$
via a set of interactions.  As the result, the mass of $S$ appears in one,
two or even three loops.

Let us consider the following K\"ahler potential:
\be\label{noscale2} K = -3 \ln(T + T^* - Z_aZ_a^*) + C_iC_i^* \;, \ee
where $T$ is the moduli field appearing in the underlying superstring
theory, $Z_a$ and $C_i$  are the matter  superfields which have the no-scale
kinetic term ($Z$--sector) and the minimal kinetic term ($C$--sector)
respectively.
The corresponding scalar potential at the Planck scale reads,
\be\label{soft2} V = |W_i|^2 + \{m_0C_iW_i + \mbox{h.c.}\} + m_0^2|C_i|^2 +
|W_a|^2 \;, \ee
where $m_0 = {\cal O}(m_{3/2})$.
The tree-level masses of the fermionic components of the fields $Z_a$ are
determined by the global supersymmetric results.
Therefore, if the  singlet fields triggering $U(1)_G$ breaking are
in the $Z$--sector, the QGF will be massless at tree level \cite{yam}.
The QGF will acquire the mass through the interactions with fields $C_i$
having minimal kinetic terms, and consequently, usual soft SUSY breaking
terms.  Moreover, $S$ (or $\s$, $\s'$) may not couple to $C_i$ directly.
It can interact with $C_i$ via couplings with some other fields $Z_a$
having no-scale kinetic terms.  In this case $S$ will get the mass in two
or larger number of loops.

Let us consider realizations of this idea in the context of the
seesaw mechanism, when $\s,\s'$ couple with right handed (RH) neutrinos $N$.
Let us introduce the following terms in the superpotential:
\be\label{seesaw}
W = {m^D \over v_2} L N H_2 + \frac{M}{f_{G}} N N \s \;,  \ee
where we have omitted the generation indices.
The first term in Eq.~\refs{seesaw} produces  the Dirac masses of neutrinos,
whereas the second one gives the Majorana masses of RH neutrino components.
The scale $f_{G} \sim 10^{10} - 10^{12} \GeV$ generates
$M \sim 10^{10}-10^{11}$ GeV required by the HDM and atmospheric neutrinos.

(i) Suppose that only $\s, \s',y$ superfields belong to the
$Z$--sector, whereas all other superfields have minimal kinetic
terms: $N, H_2, L \in C$.  Then SUSY breaking induces the soft term
\be A_N \frac{M}{f_{G}} \tilde{N} \tilde{N} \s \ee
which generates the mass of QGF in one loop (Fig.~1):
\be \label{radmass}
m_S \simeq {1\over16\pi^2} \left( M \over f_{G} \right)^2 A_N \;.
\ee
This mechanism is similar to that of the axino mass generation
by coupling of $S$ with heavy quarks \cite{yam,moh}.
For $A_N \sim {\cal O}(m_{3/2})$ and $(M/f_{G}) \sim 10^{-3}$,
$m_S$ is in the keV range.

(ii) Let us suppose that not only $\s,\s',y$ but also $N$ have the no-scale
kinetic terms.  In this case $A_N=0$ at tree level, but non-zero $A_N$
will be generated in one loop (see Fig.~2) by the soft breaking term
related to usual Yukawa interaction $LNH_2$:
$A_D m^D \tilde{L} \tilde{N} H_2$,
and by the quartic coupling $\s \tilde{N} \tilde{L}^* H_2^*$ which follows
from $|W_N|^2$ term of the supersymmetric scalar potential.
As the result  one has
\be \label{radan}
A_N \sim {1\over 16\pi^2} \left( m^D \over v_2 \right)^2 A_D \;. \ee
Correspondingly, $m_S$ appears in two loops (Fig.~2). Combining
Eqs.~\refs{radmass} and \refs{radan} we get the estimation of $m_S$:
\be \label{radmass2}
m_S \simeq {1\over (16\pi^2)^2} {A_D M^3 \over v_2^2 f_{G}^2} m_{\n} \;.
\ee
Here $m_\n = (m^D)^2/M$.
For the HDM mass scale $m_\n \simeq 3$ eV, $A_D \simeq v_2 \simeq 100$ GeV
and $f_G \simeq 10^{12}$ GeV it follows from Eq.~\refs{radmass2} that  $m_S
\simeq 3\cdot 10^{-3}$ eV can be achieved if the mass of RH component is
$M \simeq 10^9$ GeV.

In this version of model the left and right neutrino components have
different kinetic terms which may look unnatural.

(iii) Finally we consider the case where all chiral superfields belong
to the $Z$--sector. This so-called strict no-scale model \cite{nath,nano}
has only one seed of SUSY breakdown ({\it i.e.} gaugino mass).
In this case $A_D = 0$ at tree level and non-zero $A_D$
is generated in one loop by gaugino exchange.
Correspondingly, $m_S$ appears in three loops (Fig.~3) and
its estimation can be written as
\be \label{radmass3}
m_S \simeq {\a_2\over (4\pi)^5} {m_{1/2} M^3 \over v_2^2 f_{G}^2} m_{\n}
\;. \ee
Here $\a_2$ and  $m_{1/2}$ are the $SU(2)$ fine structure constant and
gaugino mass respectively.
For $m_\n \simeq 3$ eV, $m_{1/2} \simeq v_2 \simeq 100$ GeV, and $f_G \simeq
10^{12}$ GeV, one gets from Eq.~\refs{radmass3}
$m_S \simeq 3 \cdot 10^{-3}$ eV with a value of $M \simeq 10^{10}$ GeV.

A contribution to the mass of the QGF can follow also from
interactions, $W_{mixing}$, which mix $S$ with usual neutrinos (section 3).

\section{Neutrino-QGF mixing}

We now discuss possible ways which lead to mixing of the QGF
with neutrinos. Such a mixing can occur only in the presence of
either explicit or spontaneous violation of the $R$ parity
conventionally imposed in the MSSM \cite{hall}. Indeed, the Higgs field
which breaks $U(1)_G$ may belong either to $R$ even or odd
superfield depending upon the nature of the $U(1)_G$. If it
belongs to $R$ even ({\em i.e.} Higgs like) superfield then the
corresponding QGF is $R$ odd and its mixing with neutrinos
implies the $R$-violation. In contrast, if the QGF is $R$ even,
e.g. similar to the right-handed neutrino,  then its scalar
partner is $R$ odd and the $R$ symmetry gets broken together with the
$U(1)_G$ symmetry. The first alternative is realized when the
$U(1)_G$ is identified with the PQ symmetry. On the other hand,
the lepton number symmetry containing right-handed neutrino like superfield
would provide an example of the second alternative. We discuss
both these cases in turn.

\bigskip

{\it 1.\ PQ symmetry}.
The supersymmetric theories with Peccei-Quinn symmetry may contain a
term \be\label{hphi} \lambda H_1H_2\s, \ee
with $\s$ being a superfield transforming non-trivially under the PQ
symmetry.  If the axionic superfield, $S$, predominantly consists of the
field $\s$, the vacuum expectation value (VEV)
$\lag\s\rag\sim f_{PQ}$ would be large $\sim 10^{10}-10^{12} \GeV$.
Since this VEV generates the parameter $\mu=\lambda \lag\s\rag$ of the MSSM
through the interaction \refs{hphi}, one would need to fine tune $\lambda$ in
order to
understand the smallness of $\mu$. The coupling of axionic supermultiplet
$S$ to Higgs superfield is then given by
\be\label{hs1} W_{HS}=\frac{\mu}{f_{PQ}}H_1H_2S \;.\ee
The smallness of $\mu$ can be understood if $\s$ couples to Higgs through
non-renormalizable term~\cite{mu}
\be
\label{nr} \lambda H_1H_2\frac{\s^2}{M_P},
\ee
where $M_P$ is the Planck scale mass.
In this case, $\mu= \lambda\frac{\lag\s\rag^2}{M_P}$ is naturally
about the weak scale. Since $f_{PQ}\simeq \lag\s\rag$,
the axionic coupling following from Eq.~\refs{nr} can be written as
\be\label{hs3} W_{HS}=2 \frac{\mu}{f_{PQ}}H_1H_2S  \;. \ee
Alternatively, the $\s$ may acquire a small VEV $\sim \gr$
and the scale of the PQ symmetry
may be set by some other field which would predominantly
contain the axionic multiplet \cite{chun3}. The $\mu$-parameter is naturally
of the order $m_{3/2}$ in this case. As long as the field $\s$ transforms
non-trivially under PQ symmetry, it will contain a small
admixture $\sim \lag \s \rag/f_{PQ}$ of the axionic field $S$.
The interaction in Eq.~(\ref{hphi}) results in
the following coupling
\be\label{hs2} W_{HS}\sim c_\m\frac{\mu}{f_{PQ}}H_1H_2S \;,\ee
$c_\m$ being ${\cal O}(1)$.

It follows from Eqs.~(\ref{hs1},\ref{hs3},\ref{hs2})
that the axionic coupling to the
Higgs superfield is insensitive to mechanism of
implementation of the PQ symmetry. We can therefore consider the
following generic mixing term
\be\label{mix1} W_{mixing}=c_\m\frac{\mu}{f_{PQ}}H_1H_2S +
\m H_1H_2+\epsilon LH_2  \;.\ee
Here we also have included the explicit $R$ violating coupling $LH_2$.
The superpotential \refs{mix1} leads to the following mass matrix
in the basis $(\nu,S,h_1,h_2)$:
\be \label{matrix1}
\left( \ba{cccc} 0&0&0&\epsilon\\
0&m^0_S&c\mu v\sin\beta/f_{PQ}&c\mu v\cos\beta/f_{PQ}\\
0&c\mu v\sin\beta/f_{PQ}&0&\mu\\
\epsilon&c\mu v\cos\beta/f_{PQ}&\mu&0\\
\ea \right)\;, \ee
where $v \equiv \sqrt{v_1^2 + v_2^2}$ is the weak scale, $\tan\b \equiv
v_2/v_1$ and $v_{1,2}$ are the VEV's of $H_{1,2}$.
In matrix \refs{matrix1} we have included also the direct
axino mass $m^0_S$ that can be generated by the mechanisms of section 2.
We have neglected the contribution from the interactions with the gauginos
in Eq.~\refs{matrix1}.  In general gauginos mix with Higgsino through
$v_{1,2}$. This mixing will not change the qualitative results which
follow from Eq.~\refs{matrix1}. Moreover, the mixing can be small if the
gaugino mass is chosen much larger than the $\m$-parameter.
Gauginos will also
mix with neutrinos through the VEV of sneutrino field which may
arise due to the presence of the $\e$ coupling in Eq.~\refs{mix1}
and soft SUSY breaking terms. This mixing generates \cite{hallasj}
neutrino mass of order $g^2\lag\tilde{\n}\rag^2/m_{1/2}$
($g$ is the $SU(2)$ coupling constant).
For $m_{1/2}> 100 \GeV$ and $\lag\tilde{\n}\rag < 10$ keV, this
contribution is much smaller than $m^0_S \sim 10^{-3}$ eV  which
can result from the radiative corrections.

Block diagonalization of the matrix \refs{matrix1}
leads to the  following
effective mass matrix for the neutrino and the axino, $(\n,S)$:
\be\label{matrix2}
\left( \ba{cc}
0&-c\epsilon v\sin\beta/ f_{PQ}\\
-c\epsilon v\sin\beta/ f_{PQ}&m_S^0
- c^2\m v^2 \sin2\beta / f_{PQ}^2\\
\ea \right) \;. \ee
If $m_S^0 = 0$ in Eq.~\refs{matrix2}, the QGF mass,
$m_S = (2 - 3)\cdot 10^{-3} \eV$
can be obtained for  the marginally allowed value of the PQ scale:
\be
f_{PQ} \approx v \sqrt{\frac{\mu \sin 2\beta}{m_S}} \ler 4 \cdot10^9 \GeV
\;.  \ee
In this case, however, axions cannot provide the cold dark matter of the
Universe.  Note that the lightest supersymmetric particles
cannot be cold dark matter either because of their instability due to
the $R$-parity violation or due to their decay into the lighter axino.
For $f_{PQ} > 10^{10}$ GeV the QGF mass generated via $\mu$-term
is too small for the MSW solution. For $f_{PQ} \sim 10^{11}$ GeV,
$m_S  \approx 10^{-5} \eV$ is in the region of ``just-so" solution
of the solar neutrino problem.
The axion can however serve as cold dark matter provided
$f_{PQ} \sim 10^{12}$ GeV. In this case, the seesaw contribution to $m_S$
is very small and one needs a non-vanishing mass $m_S^0$.

If $m_S^0$ is the dominant contribution to the mass of $S$, $m_S \simeq
m_S^0$, one obtains from Eq.~(\ref{matrix2}) for the  $\nu-S$ mixing
\be
\label{ts1} \tan \theta_{\n s}\sim \frac{c_\m \e v \sin \beta}
{m_S^0 f_{PQ}}\;.
\ee
Then the desired value, $\tan \theta_{\n s} \sim (2 - 6)\cdot 10^{-2} \eV$
\refs{parameters}, can be obtained if  the $R$ parity breaking parameter
$\e$  equals
\be \label {epsis}
\e = \frac{m_S^0 f_{PQ} \tan \theta_{\n s}}   {c_\m  v \sin\beta}
\approx (2 - 6)\cdot 10^{-16} \frac{f_{PQ}}{\sin \beta} \;.  \ee
For $f_{PQ} \sim 10^{12}$ GeV one has $\e \sim 0.1$ MeV. In general,
the appropriate range of $\e$ is $(10^{-3} - 10) \MeV$.
It can be generated as a radiative correction:
$\e \sim h^2 m_{3/2}/16 \pi^2$.
Alternatively, $\e$ may arise through the coupling of the product $LH_2$
to some fields carrying non zero lepton number. In this case
the required smallness of $\e$ may be understood in analogy
with that of $\m$-parameter.
\bigskip

{\it 2.\ Lepton number symmetry.} Let us identify $U(1)_G$ with the
lepton number symmetry. Unlike in the previous case, it is possible
now to couple the QGF directly to neutrino through the term
\be \label{lh2} h LH_2 \s \; . \ee
This is analogous to Eq.~\refs{hphi} but now the scalar
component of $\s$ is
$R$ odd and  its VEV breaks $R$ parity. Electroweak symmetry breaking
$v_2\neq 0$ leads through the term (\ref{lh2}) to the  direct coupling
between QGF and neutrino.
Note that $\s$ is similar to the RH neutrino components.  Just as the
interaction in Eq.~(\ref{hphi}) generates the $\m$, the interaction
\refs{lh2} generates the parameter $\e$. Thus it is possible to correlate
the origin of $\e$ to the breaking of lepton number symmetry.
The smallness of $\e$ may be due to  (i) fine tuning of $h$  or
(ii) smallness of the VEV of $\s$ or due to
(iii) occurrence of the non-renormalizable coupling
analogous to that in Eq.~(\ref{nr}). All these possibilities  lead to the
following effective coupling  of $\nu$ to QGF:
\be\label{mix2} W_{mixing}=c_\e\frac{\e}{f_{L}}LH_2S +
\epsilon LH_2 \;,\ee
where $f_L$ denotes the scale associated with the spontaneous
breaking of the lepton number symmetry and $c_\e$ is a parameter
of order unity.
The mass matrix generated by Eq.~\refs{mix2} is
\be \label{matrix4} \left( \ba{cc} 0&c_\e \e v\sin\beta/ f_{L}\\
c_\e \e v\sin\beta/ f_{L} & m^0_S \\ \ea \right) \;. \ee
and the desired $\n_e-S$ mixing can be obtained for $\e \simeq 0.1$ MeV and
$f_L \sim 10^{12}$ GeV.

Let us give an example of models which leads to the mixing term of
Eq.~\refs{mix2}. Consider the $U(1)_L$ charge assignments
(1,$-1$,$-3$) for the fields $(\s,\s',L)$ respectively.
All other fields are taken neutral.
The relevant part for the $U(1)_G$ invariant superpotential is
given as follows:
\be \label{model1}
 W =  \l (\s\s' - f_{L}^2)y + {\d_\e \over M_P^2} L H_2 \s^3  \;,\ee
where the first term breaks the lepton  symmetry and generates
majoron  supermultiplet of Eq.~(\ref{qgf}).
The second  term in Eq.~\refs{model1} generates the effective
interaction displayed in Eq.~\refs{mix2}
with $c_\e={3\over\sqrt{2}}$ and  $\e\sim {\d_\e \over M_P^2} f_{L}^3$.
Thus specific choice for the lepton charges
allows one to correlate $\e$ to the scale $f_L$.
In particular, for $\d_\e \sim 0.1$ and $f_L \simeq 10^{12}$ GeV,
one has $\e \sim 1 \MeV$.
\bigskip

{\it 3.\ PQ as the lepton number symmetry.}
If both Higgs and leptons transform non-trivially under the $U(1)_G$
symmetry then the latter can play a dual role of the PQ symmetry and the
lepton number symmetry as in Ref.~\cite{lpy}.
In this case one can correlate
the origin of $\e$ and $\m$ to the same symmetry breaking scale $f_{PQ}$.
The neutrino coupling to QGF is given by
the combination of Eqs.~(\ref{mix1}) and (\ref{mix2}):
\bea\label{mix3} W_{mixing}&=&\m H_1H_2+\epsilon LH_2 \\ \nonumber
&+& c_{\m}\frac{\mu}{f_{PQ}}H_1H_2S +
c_{\e}\frac{\e}{f_{PQ}}LH_2S \;. \eea
This $W_{mixing}$ generates the following effective mass matrix for $\n$
and $S$ which is the combination of Eq.~\refs{matrix2} and
Eq.~\refs{matrix4}:
\be\label{matrix3}
\left( \ba{cc}
0&(c_\e-c_\m) \e v\sin\beta/ f_{PQ}\\
(c_\e-c_\m) \e v\sin\beta/ f_{PQ}
&m_S^0- c_{\m}^2\m v^2 \sin2\beta / f_{PQ}^2\\
\ea \right) \;. \ee
According to Eq.~\refs{matrix3} the $\n-S$ mixing angle $\theta_{\n s}$
is determined by
\be\label{ts2} \tan \theta_{\n s}\sim \frac{(c_\m-c_\e) \e v\sin\beta}
{m_S^0 f_{PQ}- c_{\m}^2\m v^2 \sin2\beta / f_{PQ}} \;. \ee

The $G$-charge prescription ($-1$,$-1$, 1,$-1$,$-2$) for
($H_1$, $H_2$, $\s$, $\s'$, $L$) permits
the following $U(1)_{G}$ invariant superpotential:
\be \label{model2}
W =  \l (\s\s' - f_{PQ}^2)y + {\d_\m \over M_P} H_1 H_2 \s^2
+{\d_\e \over M_P^2} L H_2 \s^3 \;.  \ee
It gives the terms displayed in Eq.~(\ref{mix3}) with
$c_\e={3\over\sqrt{2}},c_\m=\sqrt{2}$.

\section{Model}

Let us  put together the basic ingredients discussed in section 2
and 3 into a model  which simultaneously explains the
solar, atmospheric and the dark matter problems.
In principle the sterile state, like axino, could mix with any of the
neutrinos
but the possibility of the $\ne-S$ mixing which solves the
solar neutrino problem seems most preferred phenomenologically.
The required range of the   $\ne-S$ mixing and $S$ mass is given in
Eq.~\refs{parameters}.
The alternative possibility of $\nm-S$ mixing accounting for the
atmospheric neutrino deficit conflicts
with the cosmological bound coming from the nucleosynthesis.

Let us consider the model with $U(1)_G = U(1)_{PQ}$ broken at $f_{PQ} \sim
10^{12}$ GeV  in which the mass of QGF is generated in two or three loops
via the interaction with the RH neutrino components \refs{seesaw}
and the mixing is induced by the $L_e$-coupling described by the
superpotential \refs{model2}.
To suppress the mixing of $S$ with $\n_{\m,\t}$ and to get pseudo-Dirac
structure for $\n_\m-\n_\t$ system  (needed to explain simultaneously the HDM
and the atmospheric neutrino problem), we suggest that $U(1)_G$ is generation
dependent \footnote{One can introduce for this an additional
horizontal symmetry, suggesting that $U(1)_G$ is generation blind.}.
Consider, for example, the following prescription of $U(1)_G$ charges:
$$ \ba{cccccccccc}
  H_1 & H_2 & \s & \s' & L_e &L_{\mu}&L_{\tau}&N_e&N_{\mu}&N_{\tau}\\
  -1& -1 & 1 & -1 & -2 &-1/2&3/2&0&3/2&-1/2 \ea \;.$$
This choice gives rise to the desired phenomenological results. Specifically,
\begin{itemize}
\item
The mixing angle \refs{ts2} following from the superpotential \refs{model2}
can fall in the required range \refs{parameters} if $\e\sim 1 \MeV$ and
$f_{PQ}\sim 10^{12}\GeV$.
\item The above assignments lead to the following superpotential
in the  $\mu-\tau$ sector:
\be\label{mutau}
W = \sum_{\a=\m,\t} m^D_\a L_\a N_\a H_2
+ \frac{M_\t}{f_{PQ}} N_\t N_\t \sigma
+ \frac{M_{\m\t}}{f_{PQ}} N_\m N_\t \sigma' \,.
\ee
These couplings generate the axino mass $m^0_S$ in the MSW range
as discussed in section~2.
\item The superpotential \refs{mutau} leads to the mass matrix
in $(\n_\m,\n_\t,N_\m,N_\t)$
basis:
\begin{equation} \label{mm2}
{\cal M}=\left(
\begin{array}{cccc}
0&0&m^D_\m&0\\
0&0&0&m^D_\t\\
m^D_\m&0&0&M_{\m\t}\\
0&m^D_\t&M_{\m\t}&M_{\t}\\
\end{array}  \right ) \;.
\end{equation}
The above mass matrix gives rise to pseudo-Dirac neutrino with a
common mass
\be m_{DM}\sim {m^D_\m m^D_\t \over M_{\m\t} } \;. \ee
This mass can be in the eV range as required for the solution of
the dark matter problem by taking the values $m^D_\m \sim 0.1$ GeV, $m^D_\t
\sim 50$ GeV and $M_{\m\t} \sim 10^9$ GeV.
The mass splitting is given by
\be \label{mtsplit}
  {\D m^2 \over m_{DM}^2} \simeq 2 \left(m^D_\m \over m^D_\t \right)
                              \left(M_\t \over M_{\m\t} \right)  \;.
\ee
Taking $\left(\frac{M_\t}{M_{\m\t}} \right) \sim 1$, one reproduces
both mixing and $\D m^2$ required to explain the atmospheric anomaly.
\end{itemize}
The charge prescription, $G(N_e)=0$, permits the bare mass term $M N_e N_e$
or the non-renormalizable term $hN_eN_e \s\s'/M_P$ which will produce
$M_e \sim 10^6-10^{18}$ GeV. The Dirac mass term is generated by high-order
non-renormalizable term: $hL_eN_eH_2\s^3/M_P^3$, and therefore, $m^D_e \sim
m_e(f_{PQ}/M_P)^3$ is negligibly small.

One can get more symmetric or regular charge prescription introducing more
singlet fields or a horizontal symmetry in addition to $U(1)_G$.

The model presented above does not contain any mixing
between $\n_e$ and $\n_{\m,\t}$.
Such mixing can be induced, for example, by adding
new Higgs field which could generate a Dirac mass term
$m_{e\tau} \nu_e N_{\tau}$. This give rise to the $\nu_e-\nu_\m$ mixing
angle $\theta_{e\mu}\sim \frac{m_{e\tau}}{m_\m}$
being in the range of sensitivity of KARMEN and LSND~\cite{lsnd}
for $m_{e\tau}\sim 30 \MeV, m_\m\sim \GeV$~\cite{paper1}.

\section{Conclusions}

Simultaneous explanation of different neutrino anomalies hints to the
existence of sterile neutrino.  We have considered a possibility that the
sterile neutrino is the quasi Goldstone fermion, which appears as the result
of spontaneous breaking of a global $U(1)_G$ symmetry in supersymmetry theory.
This global $U(1)_G$ symmetry can be identified with the PQ
symmetry, the lepton number symmetry or the  horizontal symmetry.

The mass of QGF generated by SUSY breaking can be as small as
$10^{-3}$ eV so that $\n_e \to S$ resonance conversion solves the
solar neutrino problem.  In the supergravity theories such a smallness
of $m_S$ is related to special forms of superpotential and the scale of
$U(1)_G$ breaking $f_G \ger 10^{16}$ GeV or to no-scale kinetic terms for
certain superfields. In the last case, $m_S$ is generated in two or three
loops.

The mixing of QGF with the neutrinos implies spontaneous or explicit
violation of the $R$ parity.  QGF can mix with neutrino via interaction with
Higgs multiplets (in the case of PQ symmetry) or directly via coupling with
the combination $L H_2$ (in the case of lepton number symmetry).

The $U(1)_G$-symmetry being generation dependent can simultaneously explain
the dominance
of QGF coupling with electron neutrino and pseudo-Dirac structure of
$\n_\m-\n_\t$ system needed to explain the atmospheric neutrino problem and
HDM.

The PQ breaking scale $f_{PQ}\sim 10^{10}-10^{12} \GeV$ determines
several features of the model presented here. It provides simultaneous
explanation of the parameters $\e$ and $\m$ and thus leads to small
$R$-parity  violation required in order to solve the solar neutrino problem
in our approach. It also provides the  intermediate scale
for the RH neutrino masses which is required in order to solve
the dark matter and the atmospheric neutrino problem. Finally,
it controls the magnitude of the radiatively generated mass of the QGF
and allows it to be in the range needed for the
MSW solution of the solar neutrino problem. Thus the basic scenario
presented here is able to correlate variety of phenomena.

If future solar neutrino experiments
establish that  the $\nu_e -S$ conversion is the cause of the
solar neutrino deficit then one might be seeing indirect
evidence for the PQ like symmetry or for that matter of SUSY
itself.

\bigskip

{\bf Acknowledgment:} A.S.J. wants to thank ICTP for its hospitality during
his visit.

\newpage
\begin{picture}(400,300)(-200,-30)
\put(-160,0){\line(320,0){320}}
\multiput(0,0)(0,-3.2){16}{\circle*{.1}}
\multiput(-100,0)(2,3){50}{\circle*{.1}}
\multiput(100,0)(-2,3){50}{\circle*{.1}}
\multiput(0,150)(0,3.2){16}{\circle*{.1}}

\put(0,-50){\makebox(0,0){$\times$}} \put(5,-50){$\sigma$}
\put(0,200){\makebox(0,0){$\times$}} \put(5,200){$\sigma$}
\put(5,150){$A_N$}
\put(-160,7){\makebox(0,0){$S$}}
\put(160,7){\makebox(0,0){$S$}}
\put(-50,7){\makebox(0,0){$N$}}
\put(50,7){\makebox(0,0){$N$}}
\put(55,90){\makebox(0,0){$N$}}
\put(-50,90){\makebox(0,0){$N$}}
\end{picture}
\vspace{2cm}

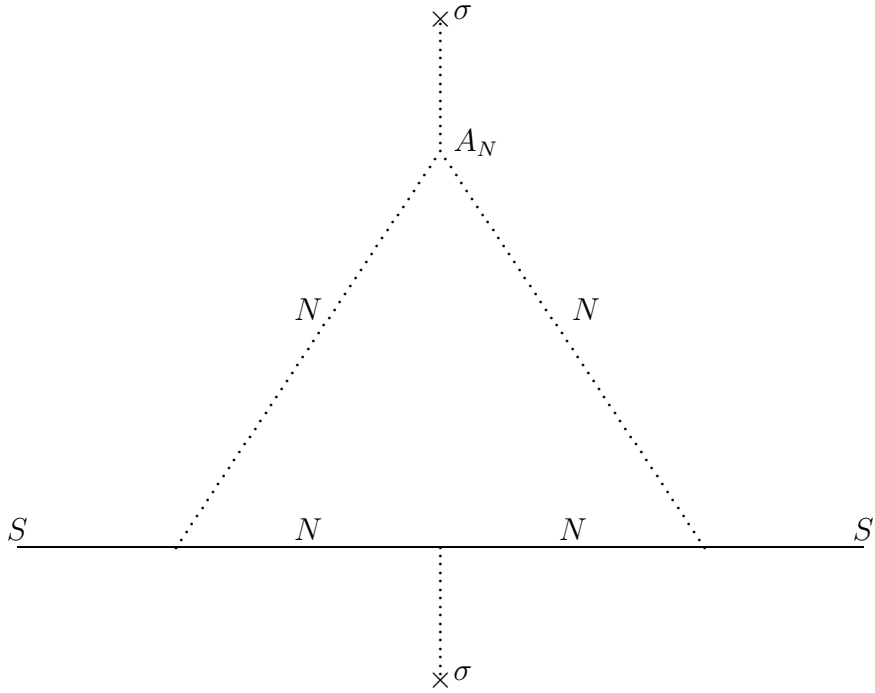
Fig.~1:  One-loop diagram for the QGF mass. The solid lines are fermions
and the dotted lines are bosons.  $A_N$ is the soft parameter of $NN\sigma$.

\newpage
\begin{picture}(400,300)(-200,-30)
\put(-160,0){\line(320,0){320}}
\multiput(0,0)(0,-2.5){20}{\circle*{.1}}
\multiput(-100,0)(0,2.5){60}{\circle*{.1}}
\multiput(100,0)(0,2.5){60}{\circle*{.1}}
\multiput(-100,150)(2,1.5){51}{\circle*{.1}}
\multiput(-100,150)(2,-1.5){51}{\circle*{.1}}
\multiput(100,150)(-2,1.5){51}{\circle*{.1}}
\multiput(100,150)(-2,-1.5){51}{\circle*{.1}}
\multiput(100,150)(2,1.5){20}{\circle*{.1}}

\put(140,180){\makebox(0,0){$\times$}} \put(135,165){$\sigma$}
\put(0,-50){\makebox(0,0){$\times$}} \put(5,-50){$\sigma$}
\put(-90,147){$A_D$}
\put(-160,7){\makebox(0,0){$S$}}
\put(160,7){\makebox(0,0){$S$}}
\put(-50,7){\makebox(0,0){$N$}}
\put(50,7){\makebox(0,0){$N$}}
\put(-15,225){\makebox(0,0){$L$}}
\put(15,75){\makebox(0,0){$H_2$}}
\put(-113,75){$N$}
\put(105,75){$N$}
\end{picture}
\vspace{2cm}

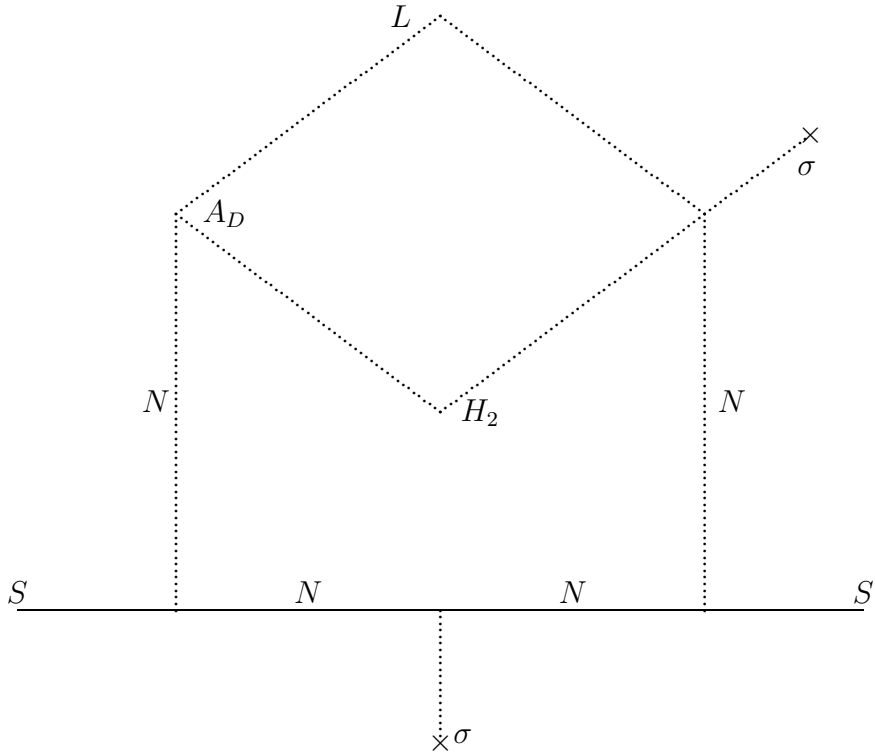
Fig.~2:  Two-loop diagram for the QGF mass. $A_D$ is the soft parameter
of $LNH_2$.

\newpage
\begin{picture}(400,300)(-200,-30)
\put(-160,0){\line(320,0){320}}
\multiput(0,0)(0,-2.5){20}{\circle*{.1}}
\multiput(-100,0)(0,2.5){60}{\circle*{.1}}
\multiput(100,0)(0,2.5){60}{\circle*{.1}}
\put(-100,150){\line(4,3){100}}
\put(-100,150){\line(4,-3){100}}
\multiput(100,150)(-2,1.5){50}{\circle*{.1}}
\multiput(100,150)(-2,-1.5){50}{\circle*{.1}}
\multiput(100,150)(2,1.5){20}{\circle*{.1}}
\put(0,75){\line(0,150){150}}

\put(0,150){\makebox(0,0){$\times$}} \put(5,150){$m_{1/2}$}
\put(140,180){\makebox(0,0){$\times$}} \put(135,165){$\sigma$}
\put(0,-50){\makebox(0,0){$\times$}} \put(5,-50){$\sigma$}
\put(-160,7){\makebox(0,0){$S$}}
\put(160,7){\makebox(0,0){$S$}}
\put(-50,7){\makebox(0,0){$N$}}
\put(50,7){\makebox(0,0){$N$}}
\put(-55,195){\makebox(0,0){$L$}}
\put(55,195){\makebox(0,0){$L$}}
\put(-55,105){\makebox(0,0){$H_2$}}
\put(55,105){\makebox(0,0){$H_2$}}
\put(-113,75){$N$}
\put(105,75){$N$}
\end{picture}
\vspace{2cm}

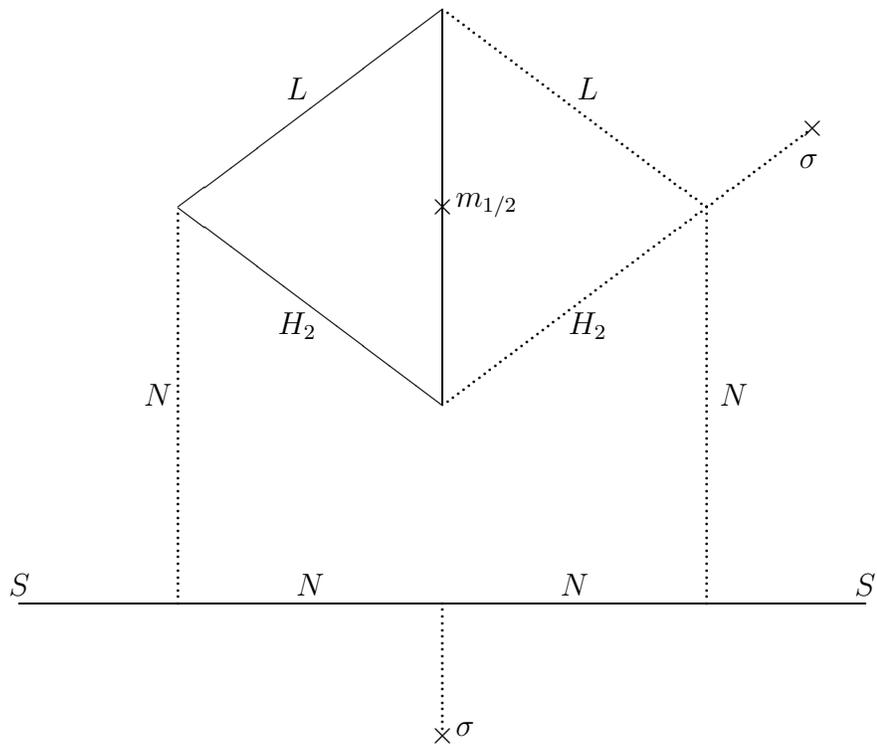
Fig.~3: Three-loop diagram for the QGF mass.
The cross with $m_{1/2}$ denotes gaugino mass insertion.

\end{document}